\documentclass{PoS}
\usepackage{epsfig}
\title{Lattice calculation of the QGP viscosities\\
- Present results and next project -}
\ShortTitle{Lattice calculation of the QGP viscosities}
\author{\speaker{Sunao Sakai}
\\

Faculty of Education, Art and Science, Yamagata University\\
Kojirakawa 1-4-12, Yamagata, Yamagata 990-8560, Japan\\ 
        E-mail: \email{sakai@e.yamagata-u.ac.jp}}

\author{Atsushi Nakamura\\

        RIISE,  Hiroshima University\\
        Kagamiyama 1-7-1, HigashiHiroshima, Hiroshima 793-8521, Japan\\
        E-mail: \email{nakamura@riise.hiroshima-u.ac.jp}}
\abstract{The shear and bulk viscosities of gluon plasma are calculated by
          accumulating a large amount of data for the Matsubara Green
          function ($G_{\beta}(t_m)$) on isotropic $24^3 \times 8$ and 
          $16^3 \times 8$ lattices.
          In the case of Iwasaki's improved action, the calculations of $G_{\beta}(t_m)$ are
          carried out on roughly 6 million configurations, while for the standard action
          the calculations are done on more than 16 million configurations.
          The shear viscosities increase roughly with
          $T^3$, and $\eta/s$ ratios are close to the KSS lower bound in
          the region where $1<T/T_c<25$. Using these data the bulk
          viscosities
          are also determined in the region where $T/T_c < 2$. They are roughly
          one order of magnitude smaller than the shear viscosities.
          Our next target is to determine the transport coefficients more
          precisely by a maximum-entropy method. For this purpose
          the most effective method may be to adopt an
          anisotropic lattice. In this report, we study the possible
          systematic error due to the deformation of the anisotropic lattice at
          short distances.  Near the critical temperature, it is found
          that the standard action suffers from a
          large deformation on the anisotropic lattice at short distances,
          while the deformation is slight for Iwasaki's improved action.
          To reduce the fluctuation of the Matsubara Green function,
          the improvement of the energy momentum tensor operator by using
          clover-type
          loops is promising.  We are
          also attempting to apply the multi-level algorithm to reduce 
          fluctuation.
          }

\FullConference{The XXV International Symposium on Lattice Field Theory\\
		 30 July-4 August 2007\\
		 Regensburg, Germany}

\begin{document}

\section{Introduction and formalism}
\indent
A new state of matter has been reported in RHIC
experiments. 
From an estimation of the temperature, it is
expected that quark gluon plasma (QGP) is realized there.
A phenomenological study of elliptic flow revealed that a new state of
matter is well explained by
the fluid with very small viscosity. Jet quenching data indicate
 that it is strongly interacting.\\
\indent
If QGP is realized as a fluid, its viscosities should be
smaller than those of usual fluids such as water and oil, because they are
 proportional to $g^{-2}$,
and the coupling constant $g$ is larger than $1/137$ even above the
transition temperature. 
The lower bound for the ratio of shear
viscosity($\eta$) to entropy(s) has recently been conjectured by 
Kovtun et al.\cite{kss}.
However, values close to transition temperature should be
calculated from QCD by fully taking into
account the nonperturbative effects. It is urgently necessary to carry out
detailed phenomenological studies of the new state of matter using these
 viscosities.\\
\indent
The transport coefficients are calculated in the framework of
 linear response theory:

\begin{eqnarray}
\eta = - \int d^{3}x' \int_{-\infty}^{t} St_{1} e^{\epsilon(t_{1}-t)} 
     \int_{-\infty}^{t_{1}} dt'<T_{12}(\vec{x},t)T_{12}(\vec{x'},t')>_{ret} 
\end{eqnarray}

\begin{eqnarray}
\frac{4}{3}\eta+\zeta = -\int d^{3}x' \int_{-\infty}^{t} dt_{1} e^{\epsilon(t_{1}-t)} 
 \int_{-\infty}^{t_{1}}dt' 
  <T_{11}(\vec{x},t)T_{11}(\vec{x'},t')>_{ret}
\end{eqnarray}

\noindent
where $\eta$ is shear viscosity, and $\zeta$ is bulk viscosity. 
$<T_{\mu \nu} T_{\rho \sigma}>_{ret}$ is the retarded Green function
of the energy momentum tensor.  In pure gauge theory it is given by
\begin{eqnarray}
T_{\mu\nu}= 2Tr[F_{\mu\sigma}F_{\nu\sigma}
-\frac{1}{4}\delta_{\mu\nu}F_{\rho\sigma}F_{\rho\sigma}],
\end{eqnarray}
and the field strength tensor $F_{\mu \nu}$ is defined by the 
plaquette operator
on a lattice:\\
$U_{\mu\nu}(x) = \exp{(ia^2 g F_{\mu \nu}(x))} $.

The transport coefficients are also expressed by the slope of the spectral
function $\rho(\omega)$ at $\omega=0$ of the corresponding retarded Green
function. The shear viscosity $\eta$ is written as 
\begin{eqnarray}
\eta = \pi \displaystyle{\lim_{\omega \rightarrow 0}}
 \frac{\rho(\omega)}{\omega}.
\end{eqnarray}
 
On a lattice, we determine $\rho(\omega)$ by the  Matsubara Green
function $G_{\beta}$ instead of the retarded Green functions themselves,
because both Green functions have the same
spectral function,\\
\begin{eqnarray}
G_{\beta}(t_m) =
-\frac{1}{\beta}\sum_{n}e^{-i\omega_n t_m} \int_{-\infty}^{\infty}
 \frac{\rho(\omega)}{i\omega_n-\omega} d\omega
= \int_{0}^{\infty}
\frac{cosh(\omega(t_m-\beta/2))}{sinh(\omega \beta/2)}
 \rho(\omega) d\omega,
\label{f-series}
\end{eqnarray}
where $\omega_{n} = 2 \pi n/\beta$ for the Matsubara Green function.
If $i\omega_n$ is replaced by $p_0+i \varepsilon$, the retarded Green function
is obtained.

However, there are still   
difficulties in the determination of $\rho(\omega)$.
One is that $G_{\beta}(t_m)$ is discrete, while $\rho(\omega)$ is
continuous. Therefore, fine resolution in the temperature
direction (simulation on large
$N_T$ lattice) is necessary for its accurate determination.
The other difficulty is that $G_{\beta}(t)$ is noisy; thus its
determination requires 
much CPU time.
Hence, we start with the smaller $N_T$ lattice, assuming a plausible form for
the spectral function $\rho(\omega)$ that fits $G_{\beta}(t_m)$ well. 
The simplest nontrivial form is\cite{karsch},
\begin{eqnarray}
\rho(\omega)
 = \frac{A}{\pi}(\frac{\gamma}{(m-\omega)^2+\gamma^2}-
 \frac{\gamma}{(m+\omega)^2+\gamma^2})
\label{rho}
\end{eqnarray}
This form is derived from a
perturbative calculation in $\varphi^4$ theory\cite{hosoya}.
\section{Numerical results from isotropic lattice}

\begin{figure}
\epsfig{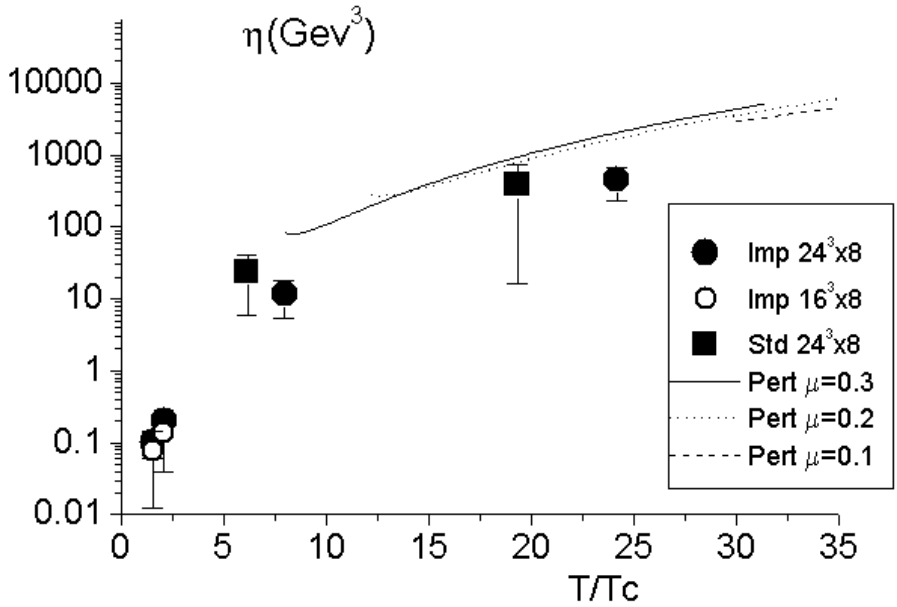}
\caption{Shear viscosity in physical units from lattice and
perturbative calculations. The circles  are results from the improved action and
squares are those from the standard action. 
The perturbative results beyond leading log approximations \cite{arnold} are shown by lines.}
\label{eta_phys}
\end{figure}

Because $\rho(\omega)$ given by Eq.(\ref{rho}) has three free parameters,
$G_{\beta}(t_m)$ should be calculated on $N_T \geq 8 $ latticels. We carry out
simulations on
$24^3 \times 8$ and $16^3 \times 8$ lattices, using Iwasaki's improved action and the
standard action, and the temperature range is $1.4 < T/T_c < 25$.
We are attempting to overcome the huge fluctuations by a large number of
 measurements.
In the case of the improved action, $G_{\beta}(t_m)$ is determined by roughly
6 million measurements, while for the standard action there are more than
16 million measurements. 
The fit of $G_{\beta}(t_m)$ is made by SALS, and
errors are estimated by the jackknife method.
Then the viscosities are  obtained by the formula
 $\eta a^3 = 4A \gamma m/(\gamma^2+m^2)^2$.\\ 
\indent
To obtain the viscosities in physical units, we need the 
lattice spacing $a(g)$. For the improved action, $a$ has been determined
for $2.2 < \beta <3.8$ by the Tsukuba group\cite{okamoto}, and for
the standard action, $a$ has been determined for
$5.58 < \beta < 6.5$ by Edward et al.\cite{edward}. 
Outside these regions, we assume a two-loop asymptotic scaling relation. \\
\indent
The results for shear viscosity $\eta$ in physical units are shown in
Fig.\ref{eta_phys}.
Because the  $\eta \times a^3$ has weak $T$ dependence,
the shear viscosity increase roughly with $T^3$ throughout the
temperature region.
We find little difference between the results from 
$24^3\times 8$ and $16^3 \times 8$ lattices. Thus, the size effect may not
be large for the lattices of these sizes. However,
 more accurate data are necessary
to determine quantitative size dependences.\\
\indent
We have also shown the perturbative results beyond the leading log
approximation\cite{arnold} in Fig.\ref{eta_phys}, where the scale factor $\mu$ in the
running coupling constant is a free parameter. 
The agreement improves when $\mu$ becomes smaller, but 
in this case the breakdown of the perturbative calculation starts at a higher 
temperature.\\
\indent
Let us proceed to the $\eta/s$ ratio, recently studied by
Kovtun et al.\cite{kss}. 
\begin{figure}
\epsfig{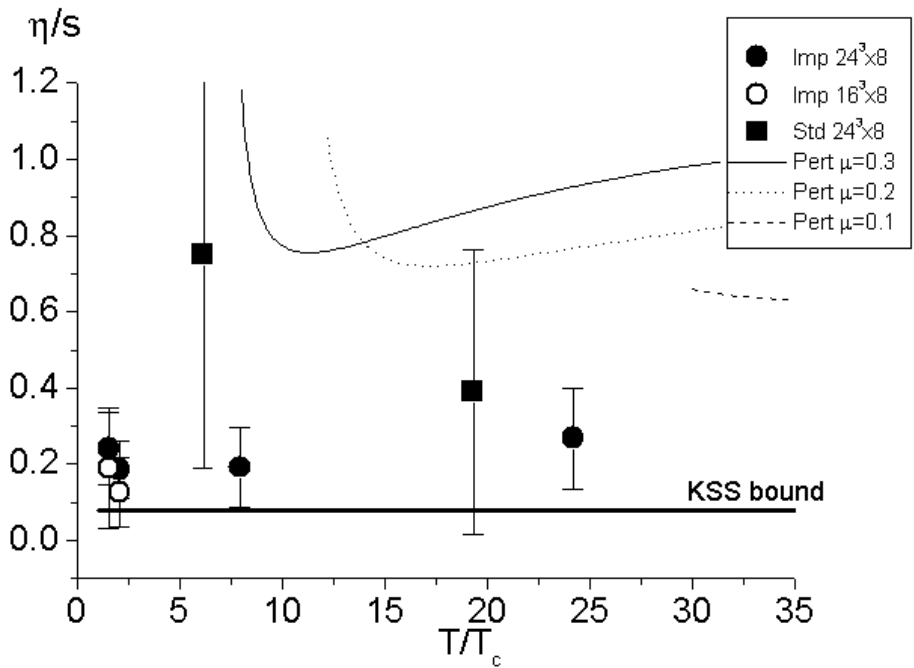}
\caption{$\eta/s$ obtained by lattice simulations (circles and squares) and
perturbative calculations (lines).
The KSS bound\cite{kss} is also shown.
}
\label{ETAVSS}
\end{figure}
The results are shown in Fig.\ref{ETAVSS}. The lattice data on 
entropies are given by Ref. \cite{okamoto,boyd}. 
The perturbative results are also shown, where we use the entropy
calculated  by the hard thermal loop approximation given in Ref.\cite{rebhan}.
Because both $\eta a^3$ and $s a^3$ have weak T dependence in the 
$T > 1.5 T_c$ region,
the $\eta/s$ ratio also has weak T dependence and the lattice results
are close to the KSS bound for $T/T_c < 25$.

\begin{figure}
\epsfig{file=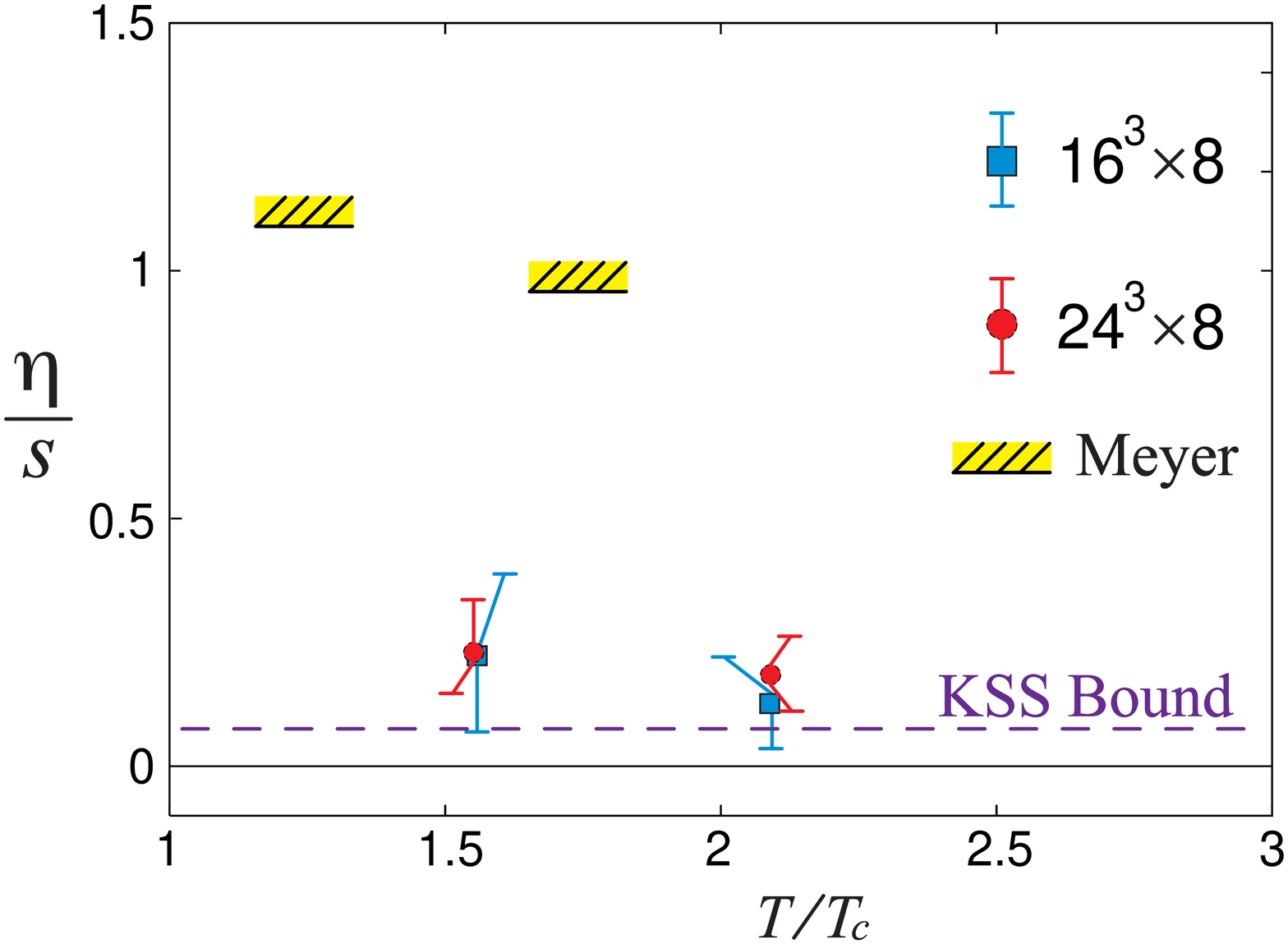, width=1.0 \textwidth}
\caption{Ratio of $\eta$/s in RHIC regions.}
\label{etaS07}
\end{figure}

In Fig.\ref{etaS07}, we show the ratio in RHIC temperature regions
together with the bounds by Meyer \cite{meyer}, who
employed a multi-level algorithm and got bounds without using an ansatz.

\begin{figure}
\epsfig{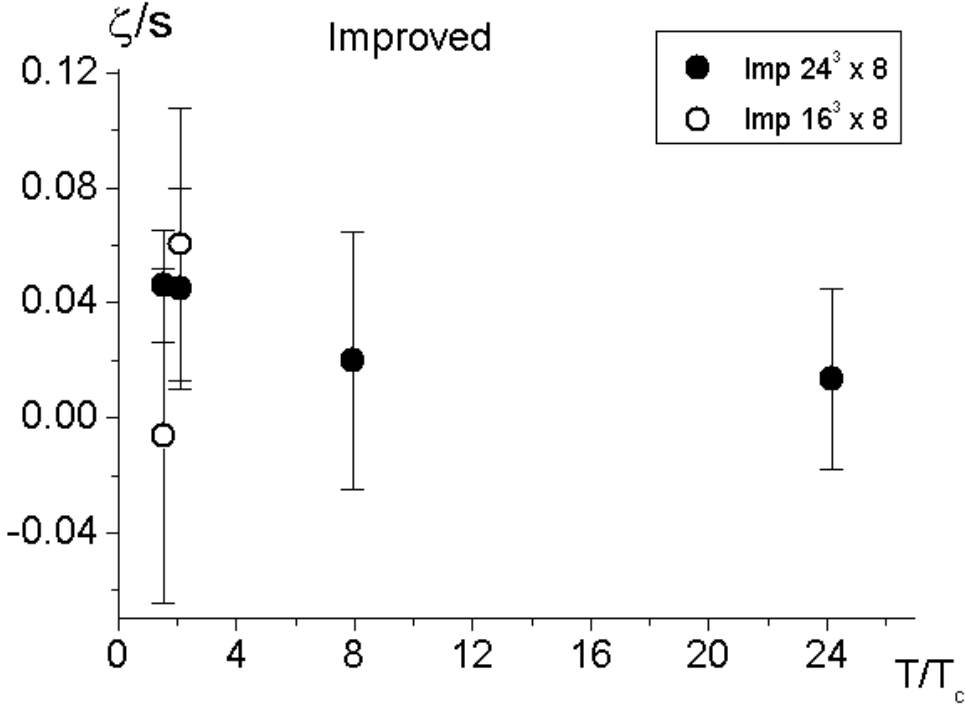}
\caption{Lattice results of $\zeta$/s ratio from the improved action }
\label{bulk}
\end{figure}

\indent
For the improved action, signals for the bulk viscosities begin to overcome
the errors, when there are about 6 million measurements.
The results are shown in Fig.\ref{bulk}. The values of bulk viscosity
still have rather large
errors, but at T close to $T_c$, their values are determined,
which are roughly one order of
magnitude smaller than the shear viscosities. Their $T$-dependence is
an interesting problem but that requires more measurements.
Recently it has been claimed that the bulk viscosity is large near the
critical temperature and decreases rapidly with T\cite{kharz}. 
Our results do not contradict this.
In the case of the standard action, the bulk viscosities still have 
large errors
that they cannot be determined.\\
\indent
It will be interesting to carry out phenomenological
studies on RHIC data taking into account these viscosities in the 
fluid model.\\

\subsection{Discussions}
$\bullet$ The renormalization factor $Z$ of the energy momentum tensor is discussed by
Meyer\cite{meyer}: $Z= 1- g^2/2(c_{\sigma}-c_{\tau})$. 
If the parametrization
of Z factor given by Ref.\cite{meyer} is used, 
the viscosities calculated by the standard
action decrease by about 30\%.
Z factor can also be written as follows: 
$Z=\partial \gamma/ \partial \xi$\cite{engles}, where $\xi$ is the
renormalized anisotropy and $\gamma$ is the bare anisotropy.  In the case
of Iwasaki's improved action, $\xi \sim \gamma$ over a wide range of
$\beta$ and $\xi$; therefore the Z factor is close to 1. If the Z factor is
taken into account, the difference between $\eta$ obtained
 from improved action and
standard action decreases.\\
\indent
$\bullet$
We have attempted to fit $G_{\beta}^{12}$ by other parametrizations of
$\rho(\omega)$ than that given in Eq.\ref{rho}.
If we apply the formula for $\rho$ proposed in Ref.\cite{Aarts}, the fit is not
satisfactory and $\rho$ does not satisfy the constraint
$\omega \rho(\omega) > 0$\cite{bellac}.
If we truncate the Taylor expansion of $\rho(\omega)$ after the lowest 3 terms,
the fit is also not satisfactory and the coefficient of $\omega$
becomes negative, which also cannot be accepted as a spectral function. 
In the three-parameter functions for $\rho$, we have not
found a $\rho(\omega)$ that fits $G_{\beta}^{12}(t_m)$ well except for Eq.\ref{rho}.

\section{Next project: toward the high-precision calculation of $\eta$}
\subsection{Simulation on anisotropic lattice}
To determine $\rho(\omega)$ without relying on any assumption, the
maximum entropy method(MEM) seems promising.  To
get a reliable result, accurate $G_{\beta}$ for a $N_{T} \sim 30$ lattice is
necessary.
For this purpose, the best method may be to adopt an anisotropic lattice. 
Before carrying out the simulation on large anisotropic lattices, we
start with a preliminary study on the possible systematic error.

\begin{figure}
\epsfig{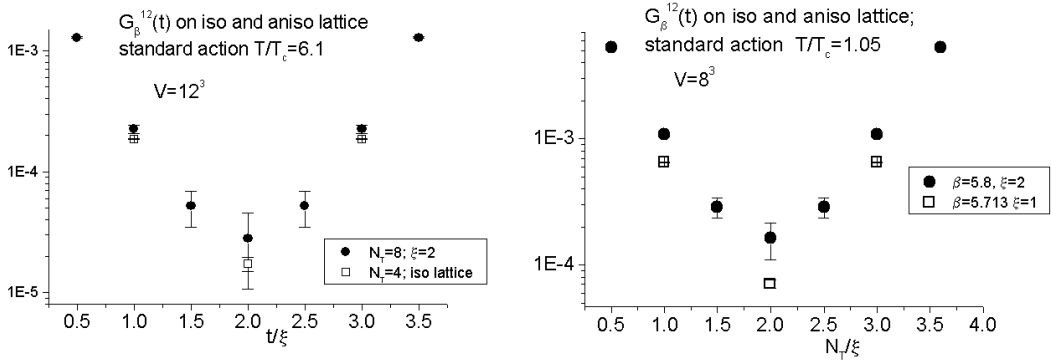}
\caption{$G_{\beta}^{12}(t_m)$ on isotropic ($N_T=4$) and $\xi=2$
 anisotropic lattice ($N_T=8$)}
\label{vol_dep}
\end{figure}

The fundamental properties of anisotropic lattices have already been
studied\cite{klassen,engles, sakai}. The anisotropy
$\xi=a_{\sigma}/a_{\tau}$  is controlled by the bare anisotropy $\gamma$
in the action. The $\gamma$ dependence of $\xi$ is expressed in terms of the ratio
$\xi/\gamma$\cite{klassen,engles,sakai}. The $\xi/\gamma$ ratio is determined by its
asymptotic plateau for $r \geq 3$, where r is the distance of the
lattice in the unit of space direction $a_{\sigma}$.  At short
distances, the $\xi/\gamma$ ratio is not equal to its asymptotic value.
Thus
$G_{\beta}(t_m)$ for small $t_m/\xi$ will suffer from a systematic error due to
deformation.\\
\indent
To study the effects of the deformation, we compare
$G_{\beta}^{12}$ for isotropic and $\xi=2$ anisotropic lattices at the
same $T/T_c$ ($a_{\sigma}$). For the standard action, $G_{\beta}^{12}$ is shown in
Fig.\ref{vol_dep}.
Large discrepancies are observed near $T_c$,
due to the deformation at short distances ($t_m/\xi \leq 2 $)
and to the difference between the Z factor in the lattices.
An increase in the
deformation is anticipated as the $\xi/\gamma$ ratio increases toward $T_c$.
On the other hand in the case of improved action, the difference 
in $G_{\beta}^{12}(t_m)$ for the lattices
is small. In this action, because
the ratio $\xi/\gamma$ is close to unity over a wide range of $\beta$ and
$\xi$, the deformation is weak and the Z factor is close to unity.

\subsection{Improvement of the energy momentum operator}
To reduce the fluctuation of $G_{\beta}$, we have attempted to use
clover-type operator for the definition
of the energy momentum tensor. It was found that the noise is strongly
suppressed, and the normalization and the $t$-dependence of
$G_{\mu \nu}(t)$ are not changed. We are also testing the effectiveness
of the
multi-level algorithm\cite{luesher} 
for reducing the fluctuations\cite{next}.

\section{Conclusion}
The shear and bulk viscosities are calculated
on isotropic $24^3\times 8$ and $16^3 \times 8$
lattices.
The T-dependence of the $\eta/s$ ratio is weak and its value is close to
KSS bound throughout the region $T/T_c < 25$.
The bulk viscosities are obtained
in the $T/T_c \leq 2$ region, and their values are one order of magnitude
smaller than those of shear viscosities.\\
\indent
The systematic error due to the deformation of lattice spacing at a short
distance is studied. For the standard action, the deformation becomes strong near
$T_c$. In the  $t_m/\xi \leq 3$ region,
the effects of deformation should be carefully controlled.\\
\indent
For the accurate determination of the spectral function $\rho$ by
MEM, improvements of the energy momentum tensor operator using 
clover-type loops and the multi-level algorithm are promising.\\

\large
\noindent
\bf Acknowledgments \\ \rm 
\normalsize
\indent
We thank R. Gupta, Y. Koma and T. Umeda for helpful discussions.
The simulations were carried out on SX5 and SX8 at RCNP. We are grateful
to the members of RCNP for their kind support.
This work was supported by Grants-in-Aid for Scientific Research
from Monbu-Kagaku-sho (The Ministry of Education, Culture, Sports,
Science and Technology) (No. 17340080).


\end{document}